\documentclass[manuscript, screen]{acmart}

\usepackage{array}
\usepackage{longtable}
\newcommand{\revised}[1]{\textcolor{black}{#1}}

\newcommand{\revadd}[1]{#1}
\newcommand{\revdel}[1]{}
\newcommand{\revchg}[2]{#2}

\newif\ifshowchanges
\showchangesfalse

\ifshowchanges
  \newcommand{\revaddII}[1]{\textcolor{blue}{#1}}
  \newcommand{\revdelII}[1]{\textcolor{red}{\sout{#1}}}
  \newcommand{\revchgII}[2]{\textcolor{red}{\sout{#1}}\,\textcolor{blue}{#2}}
\else
  \newcommand{\revaddII}[1]{#1}
  \newcommand{\revdelII}[1]{}
  \newcommand{\revchgII}[2]{#2}
\fi
\AtBeginDocument{%
  }

\acmConference[Conference acronym 'XX]{Make sure to enter the correct
  conference title from your rights confirmation emai}{June 03--05,
  2025}{Woodstock, NY}
\acmISBN{978-1-4503-XXXX-X/18/06}

\begin{document}

\title[Striking the Balance Between Efficiency and Ownership in Qualitative Data Analysis]{Not a Collaborator or a Supervisor, but an Assistant: Striking the Balance Between Efficiency and Ownership in AI-incorporated Qualitative Data Analysis}


\author{Anoushka Puranik}
\email{ap3988@rit.edu}
\affiliation{%
  \institution{Rochester Institute of Technology}
  \city{Rochester}
  \state{New York}
  \country{USA}
}

\author{Ester Chen}
\email{esterchen@mail.rit.edu}
\affiliation{%
  \institution{Rochester Institute of Technology}
  \city{Rochester}
  \state{New York}
  \country{USA}
}

\author{Roshan L Peiris}
\email{rxpics@rit.edu}
\affiliation{%
  \institution{Rochester Institute of Technology}
  \city{Rochester}
  \state{New York}
  \country{USA}
}

\author{Ha-Kyung Kong}
\email{hidy.kong@rit.edu}
\affiliation{%
  \institution{Rochester Institute of Technology}
  \city{Rochester}
  \state{New York}
  \country{USA}
}

\begin{abstract}
Qualitative research offers deep insights into human experiences, but its processes, such as coding and thematic analysis, are time-intensive and laborious. Recent advancements in qualitative data analysis (QDA) tools have introduced AI capabilities, allowing researchers to handle large datasets and automate labor-intensive tasks. However, qualitative researchers have expressed concerns about AI's lack of contextual understanding and its potential to overshadow the collaborative and interpretive nature of their work. This study investigates researchers' preferences among three degrees of delegation of AI in QDA (human-only, human-initiated, and AI-initiated coding) and explores factors influencing these preferences. Through interviews with 16 qualitative researchers, we identified efficiency, ownership, and trust as essential factors in determining the desired degree of delegation. Our findings highlight researchers' openness to AI as a supportive tool while emphasizing the importance of human oversight and transparency in automation. Based on the results, we discuss three factors of trust in AI for QDA and potential ways to strengthen collaborative efforts in QDA and decrease bias during analysis.

\end{abstract}



\maketitle

\section{Introduction}
Qualitative research is a foundational methodology used across diverse disciplines such as human-computer interaction, sociology, psychology, anthropology, and healthcare \cite{Creswell2016QualitativeInquiry, Denzin2018SageHandbook}. It provides deep insights into complex human behaviors, experiences, and perspectives, often addressing aspects that quantitative methods cannot capture \cite{Patton2014QualitativeResearch}. One of the most widely used methods within qualitative research is thematic analysis, \revchg{where researchers perform "coding" on the qualitative data}{a process for identifying, analyzing, and reporting patterns (themes) within data} \cite{braun2006using}. \revadd{In qualitative research, coding involves assigning specific, granular labels to raw data segments to create the foundational building blocks of analysis \cite{braun2006using, Saldana2013}. These codes are subsequently clustered and synthesized into themes, which are broader patterns and interpretive insights that capture significant meaning within the dataset \cite{braun2006using, Saldana2013}.} While these qualitative data analysis (QDA) methods generate rich and nuanced findings, this process is often socially intensive, frequently involving teams of researchers engaging in discussion, negotiation, and interpretation to reach shared \revchgII{understandings}{understanding} of the data \cite{kuang2022merging, gao2024collabcoder, Guest2012AppliedThematic}. As qualitative datasets grow larger and more complex, the collaborative labor involved in coding and interpreting becomes increasingly time-consuming and labor-intensive, and is susceptible to implicit bias due to the subjective nature of the process \cite{Salmons2015QualitativeOnline}. Given the extensive nature of this work, qualitative researchers often rely on QDA tools to facilitate organization and streamline \revdelII{the} workflow \cite{sotiriadou2014, welsh2002, Silver2014UsingSoftware}.  
In addition, researchers have looked into ways to incorporate artificial intelligence (AI) to support QDA \cite{gao2024collabcoder, riet2021, overney2024, feuston2021human_ai_collab}. Recently, QDA tools including NVivo and MAXQDA started automating various tasks, such as coding and theme generation, reducing the manual effort involved in analysis. AI-powered QDA tools offer the potential to streamline workflows, manage large datasets, and improve efficiency \cite{yan2024human}. 

Despite these advancements, the integration of AI into the collaborative practices of qualitative analysis was not immediately embraced by all qualitative researchers \cite{hilbolling2024, jiang2021, Neang2021Data, Jonathan2008Sharing, Joel2016Applying}. An earlier study showed that researchers enjoy the discoveries they make through manual coding due to the interpretive nature of qualitative research and resist the notion of AI fully automating this work \cite{jiang2021}. 
More recent studies show a growing openness of qualitative researchers to collaborating with AI, and suggest ways in which AI can or should be used \cite{feuston2021human_ai_collab}. Yet, while researchers have proposed AI as a novel form of collaborator \cite{gao2024collabcoder}, little is known about how people perceive this AI ``collaborator" compared to human collaborators, and how AI might affect or augment the collaborative nature of QDA.

Given the conflicting views on the integration of AI in QDA, this study explores how qualitative researchers perceive the role of AI in their collaborative analysis workflows, using Lubars and Tan's framework on human perception of task delegation to AI \cite{lubars2019delegability}. Their framework divides the degree of delegation into four categories: no AI assistance, the human leads and the AI assists, the AI leads and the human assists, and full AI automation. We adjusted the degrees to describe different types of AI involvement in the QDA process: human-only, human-initiated\revaddII{,} and AI-initiated coding. 
\begin{itemize}
    \item \textit{Human-only coding} involves human coders manually assigning codes based on their interpretation of the data, either by hand or using QDA software without any AI involvement.
    \item \textit{Human-initiated coding} refers to a process where \revdel{themes and}codes are created and applied by \revchg{a human coder and later verified using AI tools}{a human coder, grouped into themes, and the resulting codes and themes are later reviewed using AI tools.}
 
    \item \textit{AI-initiated coding}, in contrast, uses machine learning algorithms to automatically \revchg{detect}{identify} \revchg{themes}{codes} and assign \revadd{these} codes \revadd{to data segments, or cluster codes into broader themes,} with human coders reviewing the results afterward.
\end{itemize}

We omitted the full automation category given the strong resistance shown in prior work \cite{jiang2021}. Based on this framework, we examined the desired role of AI in the QDA process through interviews with 16 qualitative researchers. We investigated the topic through the lens of how their preferences align with existing human collaboration in QDA and how AI might shift or support these collaborative practices. We further explored the topic of bias in QDA and participants’ use of multimodal data (e.g., audio and video), as it represents an underexplored strategy for mitigating bias through capturing nuances like tone, pitch, and non-verbal cues \cite{craig2021engaging}. Through the study, we aimed to answer the following three research questions:

\begin{itemize}
    \item \textbf{RQ 1}. \revchg{Which degree of AI involvement (human-only, human-initiated, AI-initiated) in QDA is most preferred by qualitative researchers, and what are the factors leading to this preference?}{How do qualitative researchers’ preferences for different degrees of AI involvement in QDA (human-only, human-initiated, and AI-initiated coding) vary across different contexts, and what factors shape these preferences?} 
    \item \textbf{RQ 2}. What are qualitative researchers’ perspectives on bias in qualitative data analysis and the use of AI for mitigating bias?
    
    \item \textbf{RQ 3}. How do qualitative researchers perceive the value of multimodal data (e.g., audio, video), and how do they incorporate it into their analysis practices?
\end{itemize}

Our study showed that human-only coding is still the most preferred method, as participants were reluctant to give interpretive authority to AI. Yet, they were open to workflows where AI acts as an assistant, not collaborator or supervisor, either by generating code suggestions that humans can accept or reject, or by automating repetitive tasks. Many expressed discomfort in human-initiated coding, where AI \textit{checks} human work, as it undermined their sense of ownership of the codes and professional identity as a qualitative researcher. 
The interviews surfaced a mistrust in the involvement of AI in the QDA workflow due to a lack of contextual understanding and concerns on data confidentiality. Despite recognizing the potential of audio and video to enrich analysis, participants relied heavily on text due to time constraints and inadequate tools, highlighting an opportunity for future QDA systems to integrate multimodal analysis. Through the study, we make the following three primary contributions:

\textbf{1. Qualitative comparison of AI delegation paradigms in QDA:} We explored researchers’ preferences across three degrees of delegation for AI in the QDA workflow (human-only, human‑initiated, and AI‑initiated coding) and show how efficiency, ownership, and trust influence people's preferred degree of delegation.

\textbf{2. Design implications that position AI as an assistant rather than a collaborator:} We propose concrete CSCW designs that use AI as a method of enhancing human collaboration in QDA by distinguishing authorship and establishing trust. 

\textbf{3. Highlighting opportunities for mitigating bias, including multimodal integration:} 
We address the underutilization of multimodal data in current QDA workflows and propose AI-driven approaches for incorporating non-verbal cues to support richer and more bias-aware analysis.

\section{Related Work}

\subsection{Tools for Qualitative Data Analysis}
Historically, manual qualitative data analysis (QDA) presented significant logistical hurdles. The transcription process alone was exceptionally time-consuming; for instance, a mere 45-minute audio-recorded interview could demand approximately eight hours of a trained researcher's time, resulting in 20 to 30 pages of written text for analysis \cite{sutton2015qualitative}. This extensive manual effort in data preparation and handling was a major bottleneck in the research workflow.

When the QDA is performed collaboratively as a team, especially one that is geographically distributed or working on complex projects, a distinct set of challenges emerges \cite{jiang2021, Neang2021Data, Jonathan2008Sharing, Joel2016Applying}. These difficulties can be broadly categorized into three main areas.
\textit{First}, technical infrastructure issues pose significant barriers. These include incompatibilities between different software tools, problems with secure data sharing, challenges related to legacy data, differing file standards, and complexities reminiscent of version control problems \cite{Dylan2007Can, Alan2016Abstracts, C2020Collaborative}.  
\textit{Second}, methodological coordination presents considerable challenges. Teams often struggle with overcoming the absence of standardized analytical procedures, aligning coding schemes, and ensuring and managing inter-coder reliability \cite{jiang2021, Joel2016Applying}.
\textit{Third}, team communication and workflow management are often problematic, especially in remote settings. Difficulties in decision-making, effective knowledge sharing, and coordinating efforts across different time zones are key challenges in such contexts \cite{Zreik2022Collaborative, Dylan2007Can, Heckemann2020Working, Prikladnicki2008Conducting, Neang2021Data}.

The advent of Computer-Assisted Qualitative Data Analysis Software (CAQDAS) marked a significant shift in the QDA landscape, addressing many of the initial manual challenges \cite{evers2011past, gilbert2014tools}. Leading CAQDAS tools like NVivo \cite{nvivo} and ATLAS.ti \cite{atlas} began to streamline the research process by offering features such as automated transcription and more efficient management of large datasets \cite{evers2011past}. These tools considerably reduced the time required for thematic analysis by allowing researchers to sort data by codes and summarize coded segments across multiple interviews. While some CAQDAS packages offered assistance in theory development and advanced search functionalities, they fell short in areas such as recognizing synonyms, which could hinder the extraction of deeper insights \cite{bringer2006, welsh2002}. Other tools focused on strengths like automated concept mapping, robust data storage, and flexible management capabilities \cite{sotiriadou2014, soratto2020}, illustrating the ongoing technological evolution within the field.

Despite the advancements brought by CAQDAS, \revchg{some challenges persisted paving the way for the integration of Artificial Intelligence (AI) into the QDA process.}{limitations persisted, particularly in tasks requiring high-volume pattern recognition and time-intensive data sorting, which are uniquely suited for Artificial Intelligence (AI).} Marathe and Toyama \cite{marathe2018semi} identified that many coding practices across various disciplines could potentially be automated. The application of computational methods in qualitative research promised enhanced efficiency and a potential reduction in bias and errors \cite{abramson2018promises}. Consequently, the research community began to explore automation processes \cite{wiedemann2013opening, blank2014big}, and some CAQDAS tools incorporated features like ``Automated Insights'' in NVivo, which leverage topic modeling \cite{automatedInsights}.
Further research into semi-automated systems such as Cody, which integrates user-defined rules with supervised machine learning, demonstrated improvements in coding efficiency and consistency \cite{riet2021}. Studies by Overney et al. (2024) \cite{overney2024} and van der Riet et al. (2021) \cite{riet2021} have further highlighted AI's capacity to enhance coding quality and inter-rater reliability through AI-assisted coding.

\revadd{While these earlier works focused on augmenting individual efficiency, recent HCI and CSCW research has shifted toward exploring how AI could be integrated to enhance the \textit{collaborative} aspect across the full workflow of qualitative inquiry from dynamic data collection, collaborative coding, to interpretive partnership.} \revchg{Jiang et al. \cite{jiang2021} investigated how AI could augment, rather than fully automate, inductive and interpretive research workflows.}{First, in the \textit{data collection phase,} researchers are reshaping how information is acquired by collaborating with AI. Cuevas et al. \cite{Cuevas2025Collecting} demonstrated that LLM-based interviewing chatbots can act as collaborative partners in gathering data, fostering new forms of dynamic engagement between researchers and participants that was previously difficult to scale. Lei et al. \cite{Lei2025Dynamic} also explored AI-supported data collection, but through dynamic surveys where LLM is used to cluster responses in realtime and prompt survey respondents to reflect on the answer trends. Their work aimed at designing survey platforms that support both quantitative structure and collaborative interactions.}
\revadd{Second, in the \textit{coding and consensus phase,} Wang et al. \cite{Wang2025LATA} explored the feasibility of ``LATA'' (LLM-Aided Thematic Analysis), highlighting that while AI can mimic the coding process, it introduces challenges in reliability that require human oversight. Consistent with these concerns, Gao et al. \cite{Gao2023CoAIcoder} investigated the dynamics of AI-mediated collaborative coding through CoAIcoder, and found that while AI suggestions improved efficiency and agreement, they also introduced a risk of ``shallow agreement,'' where human researchers converged on the AI's suggestions rather than engaging in deep, independent interpretation.}
\revadd{Third, regarding \textit{high-level interpretation,} literature suggests a shift from automation to ``augmentation'' of serendipity.} Jiang et al. \cite{jiang2021} \revchg{investigated how AI could augment, rather than fully automate, inductive and interpretive research workflows. Their work, based on interviews with qualitative researchers, emphasized the importance of transparency, human agency, and support for serendipity in the design of AI-augmented CAQDAS tools.}{framed the AI not as a tool for automation, but as a system that honors uncertainty and supports serendipity, arguing that the goal of collaboration is to expose researchers to unexpected patterns while maintaining human agency.} \revchg{Expanding on this, Feuston and Brubaker \cite{feuston2021human_ai_collab} explored AI's potential role across various qualitative analysis stages, including data exploration, sampling, and initial coding, through storyboard scenarios. Their findings shed light on shifts in scale, abstraction, and delegation necessary for designing effective human-AI collaborative systems.}{Feuston and Brubaker \cite{feuston2021human_ai_collab} expanded on this by using storyboard scenarios to map the ``delegation'' of tasks, identifying how researchers negotiate shifts in scale and abstraction when collaborating with AI.} \revadd{Finally, moving into theme construction, Kang et al. \cite{Kang2025ThemeViz} explored how interactive systems (ThemeViz) can mediate the ``sense-making'' process, allowing researchers to actively negotiate and refine AI-generated themes rather than simply accepting them.}

\revdel{However, the integration of AI is not without its own set of challenges. Current topic models, for example, may generate low-quality topics or topics that misalign with existing domain knowledge, thereby potentially reducing the accuracy of analytical outcomes \cite{smith2018closing}. Moreover, qualitative researchers expressed reservations about the automation of thematic analysis. As Evers notes \cite{evers2018current}, qualitative research values not just patterns but also subtle differences in meaning and even the absence of certain utterances, aspects that fully automated systems might overlook. While advanced AI models like GPT-4 show potential for rapid prototyping via prompt engineering to get better results in qualitative coding, they still fall short in tasks requiring nuanced data interpretation \cite{kirsten2024}. Consistent with these concerns about AI’s limitations in nuanced interpretation, Gao et al. showed that while AI-mediated coding can improve efficiency and agreement in early stages, it may also encourage researchers to reuse existing codes and converge too quickly, leading to a ``shallow agreement'' where researchers accept the AI's first suggestion rather than critically engaging with the text to develop rich, novel interpretations.  \cite{Gao2023CoAIcoder}. Similarly, Gao et al. highlighted user reluctance to overly rely on AI for complex analyses like paragraph-level coding, observing a tendency for researchers to reuse previous codes, which can result in shallower interpretive coding.}

\revadd{Despite these collaborative advances, significant challenges remain regarding the depth of interpretation. Qualitative research values subtle differences in meaning and the absence of utterances \cite{evers2018current}, nuance that current models often miss \cite{smith2018closing, kirsten2024}. This tension between efficiency and interpretive depth remains a central hurdle in Human-AI QDA.}
Our research extends \revchg{the existing body of work on human-AI collaboration in QDA that have laid foundational groundwork by exploring researcher needs and potential AI roles as well as demonstrated enhanced efficiency of related systems \cite{jiang2021, feuston2021human_ai_collab, riet2021}. More specifically, we focus on the desired degree of delegation based on Lubars and Tan's framework \cite{lubars2019delegability} and investigate how qualitative researchers' view on human-AI collaboration in QDA differs from current human collaboration in QDA.}{this groundwork by focusing specifically on the \textit{mechanisms of delegation}. Building on Lubars and Tan's framework \cite{lubars2019delegability}, we investigate whether researchers view the ``AI partner'' similarly to a human collaborator (requiring negotiation) or a research assistant (requiring supervision), and how these views shape the desired ``collaborative contract'' in QDA workflows.}

\subsection{Bias in Qualitative Data Analysis}

\revchg{Qualitative coding involves subjective judgment calls.}{Qualitative analysis is inherently interpretive, relying on the researcher's subjectivity as a lens to make sense of data.} \revadd{However, a distinction must be drawn between \textit{subjectivity}, which is a necessary feature of qualitative inquiry and \textit{bias}, which threatens validity.} Drapeau explained that \revadd{unexamined} subjectivity may be introduced in qualitative research\revdel{ on controversial topics}, \revadd{particularly} where there is a lack of consensus on definitions, \revchg{and researchers code according to their internalized definition that may be the result of strong prejudice or personal conflicts”}{leading researchers to code based on internalized prejudices or conflicts} \cite{drapeau2002subjectivity}. In addition, researchers are often influenced by how the results may be used later on (e.g., in court, by the media) and could lead to an unconscious decision to derive conclusions based on the moral and legal implications rather than the data present \cite{drapeau2002subjectivity}. \revchg{Both factors illustrate how unexamined subjectivity can lead to cognitive biases, such as confirmation bias, where researchers select data and interpretations that fit their pre-existing beliefs or needs, especially for controversial or stigmatized research topics \cite{onwuegbuzie2007validity}.}{In HCI and CSCW, \textit{reflexivity} is the standard method for mitigating these risks, requiring researchers to critically examine their own positionality \cite{finlay2002negotiating, harrison2007paradigms}. Without this, unexamined subjectivity can harden into cognitive biases, such as confirmation bias, where researchers select data that fits pre-existing beliefs \cite{onwuegbuzie2007validity}.} \revadd{This risk is particularly acute when researchers rely solely on text transcripts; the removal or underutilization of multimodal data (e.g., audio tone, hesitation) can strip away the nuance needed to challenge a researcher's initial textual interpretation, thereby exacerbating confirmation bias.}

\revadd{While human researchers struggle with cognitive bias, AI introduces a different challenge.} Given that AI systems are trained on human-provided datasets, \revchg{some biases can inadvertently seep into AI-generated results.}{societal biases can be encoded into the model, resulting in \textit{algorithmic bias}.} Panch et al. \cite{panch2019ai_bias} discuss this form of \revdel{algorithmic} bias and stress the importance of designing AI systems to actively mitigate it. Our study builds on these discussions by exploring qualitative researchers' views on \revchg{bias in QDA, and the potential role of QDA tools in addressing these challenges.}{how these two distinct forms of bias --- human cognitive bias and AI algorithmic bias interact and can be mediated.} Prior work by Ma et al. on trust calibration (i.e., comparing the user's trust in AI with AI's actual capability) showed that to establish an adequate level of trust, we not only have to assess AI's correctness likelihood but also the user's correctness likelihood \cite{ma2023trust}. \revchg{Similarly, we investigate the interplay between humans' cognitive bias and AI's algorithmic bias to understand how to effectively balance the strengths of both.}{Similarly, rather than viewing AI as a neutral adjudicator, we investigate how QDA tools might allow researchers to leverage AI to identify their own blind spots, while simultaneously using their human contextual awareness to audit algorithmic limitations.}

\section{Methodology}

\revadd{We define "codes" and "themes" in alignment with established qualitative research practices \cite{Saldana2013, braun2006using}. \textit{Codes} are the specific, granular labels (word/short phrase) assigned directly to segments of qualitative data (e.g., text and audio transcripts) and serve as raw building blocks. \textit{Themes} represent broader, more abstract patterns or insights that emerge from clustering, categorization, or analytic reflection of individual codes. Our study investigates AI's role in QDA, with the AI assistance primarily directed at supporting the coding phase by generating suggestions for these granular codes. The subsequent process of developing broader themes remains a researcher-led interpretive task.}

\subsection{Participants}
 We recruited individuals across various departments involving qualitative research (e.g., sociology, history, anthropology, and information science) through campus flyers, email outreach, and professional networking platforms such as LinkedIn. \revadd{Inclusion criteria for eligible participants included a minimum of one year of experience in qualitative research and prior experience using a qualitative data analysis tool.} \revdel{The inclusion criteria were - 
\\1. A minimum of one year of experience in qualitative research.
\\2. Prior experience using a qualitative data analysis tool.}
\\Interviews were conducted with 16 participants, aged 23 to 40 (mean = 29.1), with qualitative analysis experience ranging from 2 to 7 years (mean = 3.9). 
\revised{Participants reported their highest degree completed or currently being pursued: 9 were pursuing or had completed a PhD, and 7 a Master’s degree.}
\revised{Eleven participants were in academia (8 PhD students and 3 MS students), and five participants were user experience researchers in industry.}
The research expertise of participants ranged from accessibility, human-machine teaming, behavioral decision-making, paralinguistic elements in captions for the Deaf and Hard-of-Hearing, early childhood education, to HIV counseling. More information is provided in Table \ref{tab:participant_info}.

\subsection{ChromaScribe: AI-based QDA Prototype}
\begin{figure}[h!]
    \centering
    \includegraphics[width=1.0\linewidth]{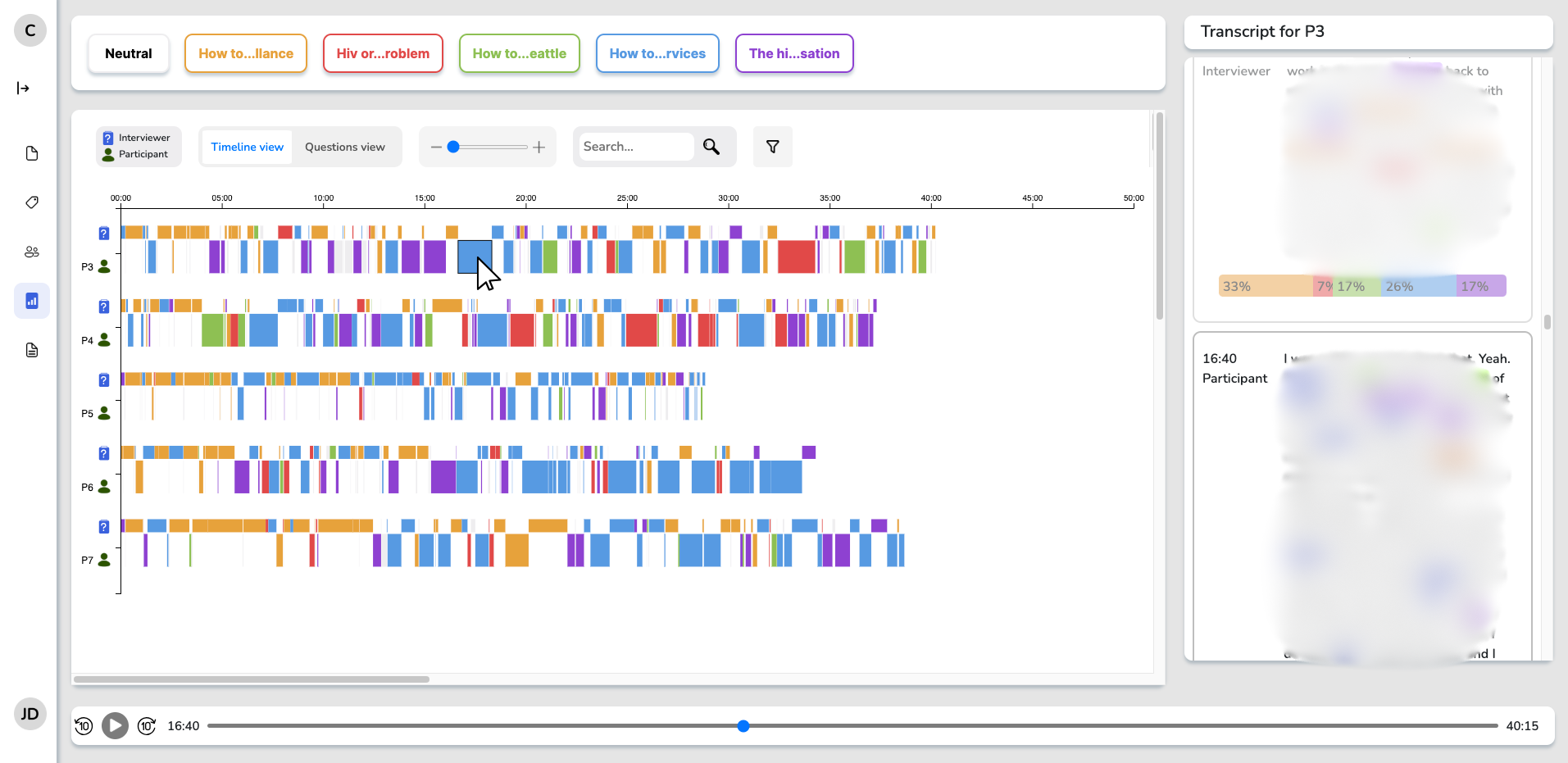}
    \caption{The main page of ChromaScribe, an AI-augmented QDA webtool prototype with a color-coded visualization panel \revchg{on the left side}{(left)} and transcription panel \revchg{on the right}{(right). A mouse cursor highlights the user selecting a blue thematic block between 15:00 and 20:00 for participant P3. Consequently, the transcription panel on the right displays the specific transcript segment corresponding to this selection}. \textit{(Note, \revchg{the transcript portions have been}{Transcript text is} blurred to preserve the confidentiality of interview data)}.}
    \label{fig:overall_vis_screenshot}
\end{figure}

To ground the discussion and give participants an idea of what AI integration in QDA might look like, we developed ChromaScribe as shown in Figure~\ref{fig:overall_vis_screenshot}, an AI-based QDA webtool prototype. \revadd{The tool was introduced as an exploratory artifact rather than a fully realized QDA system. It was not designed to adhere to a single AI-delegation type, and it was possible to use the tool for both the AI-initiated and human-initiated workflow. \revadd{This choice follows established prior work showing that early-stage prototypes are effective for exploration and reflection rather than evaluation, helping participants articulate needs, expectations, and values around emerging technologies \cite{tohidi2006getting, Gao2023CoAIcoder, feuston2021human_ai_collab}.} Thus, the prototype served as a concrete prompt to support discussion around AI-supported qualitative analysis, including perceived benefits, concerns, and feasibility.} 

\begin{figure}[h]
    \centering
    \includegraphics[width=1.0\linewidth]{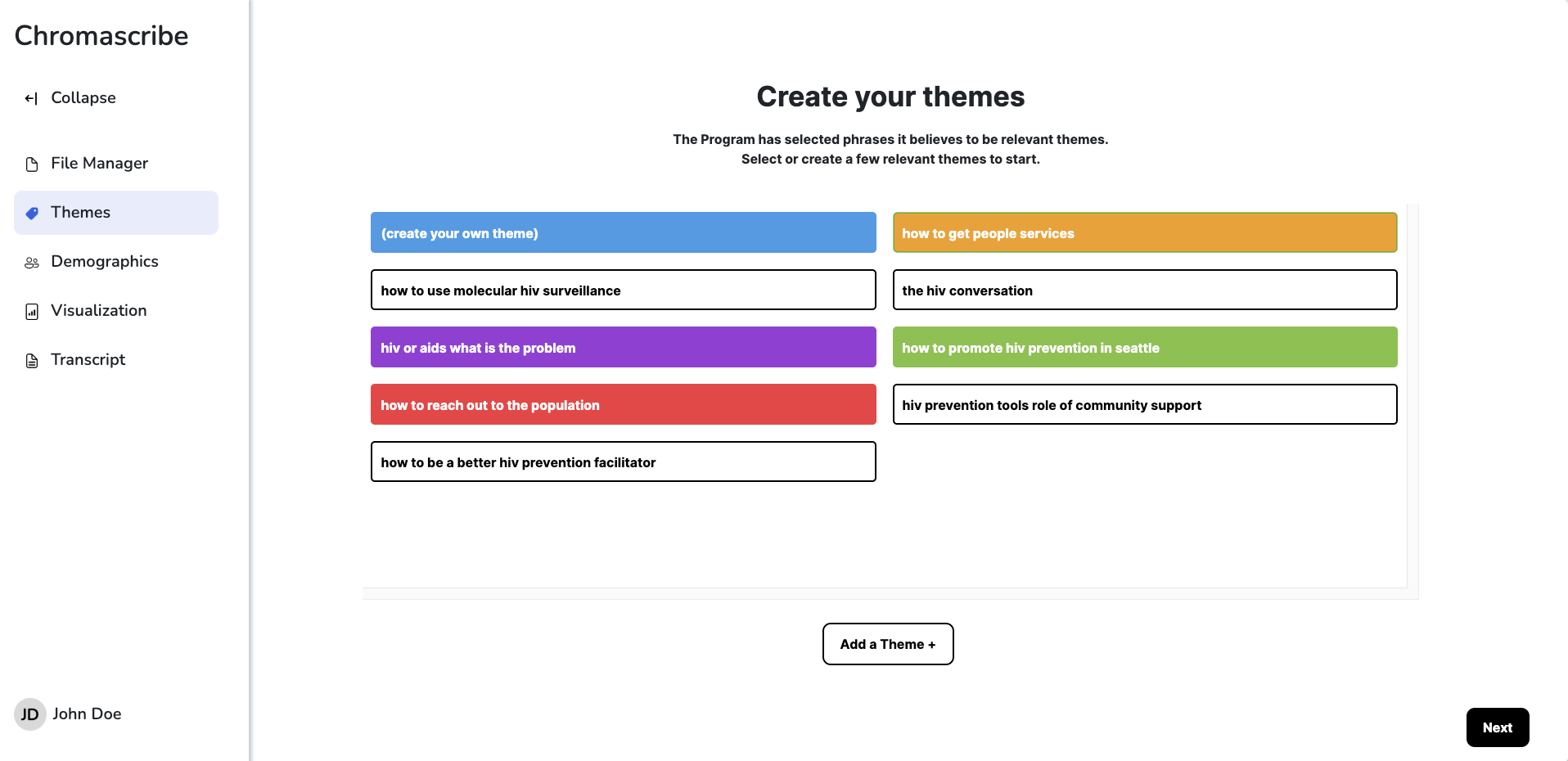}
    \caption{The theme selection page shows AI-generated themes. The user can select automated themes and/or add their own themes using the button at the bottom. Selecting a theme assigns a color to the theme. The selected themes are automatically applied to all transcripts and visualized on the main page. }
    \label{fig:Automated theme generation}
\end{figure}

\begin{figure}[h]
    \centering
    \includegraphics[width=1.0\linewidth]{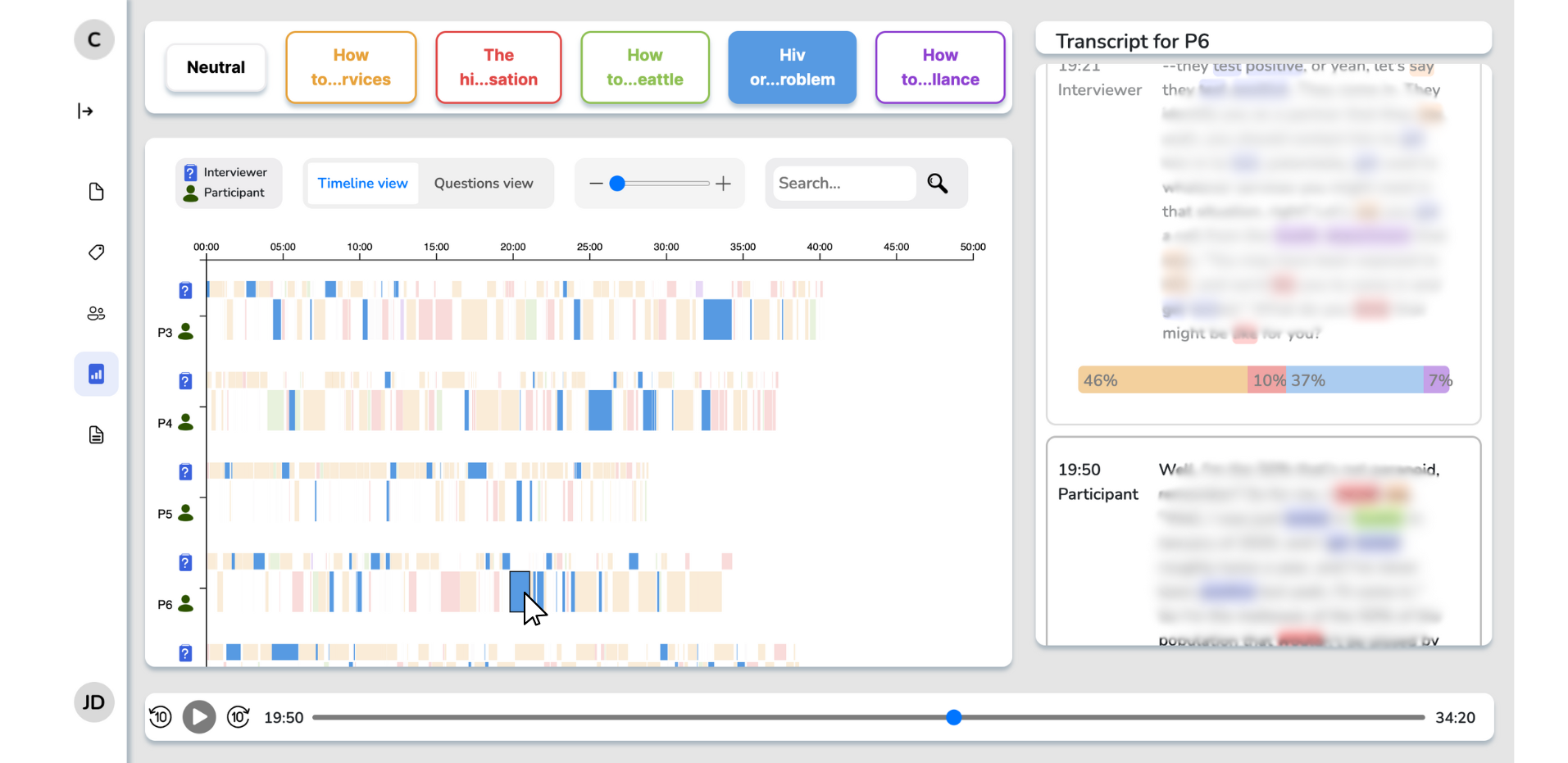}
    \caption{The visualization panel shows a timeline-based visualization of theme distributions. When a user selects a theme (such as the blue theme selected \revadd{by the mouse cursor} \revdel{here}), all transcript segments that are coded with the theme get highlighted in the visualization. \textit{(Note, the transcript portions have been blurred to preserve the confidentiality of interview data)}.}
    \label{fig:theme_viz}
\end{figure}

\revaddII{However, its inclusion of AI-generated themes could have functioned as a framing device. By exposing participants to these themes, the system may have anchored participants' mental models of the role of AI in QDA. The study findings should be interpreted within this context.} The webtool uses a locally hosted interface to ensure the data is not accessed by any third-party cloud servers. Access to the prototype was restricted through a password-protected login, and credentials were shared only with study participants for the duration of the study. To further ensure confidentiality, the access password was changed after each study session. Upon logging in, participants encountered a pre-loaded dataset, which consisted of \revadd{anonymized} interview transcripts and corresponding audio recordings from a study on HIV molecular surveillance involving healthcare workers and individuals from affected communities. The tool automatically identifies and color-codes themes based on transcripts and audio files, generating a set of preliminary themes using natural language processing, which are then presented to users for review. \revadd{We selected established, non-generative natural language processing (NLP) models over opaque Large Language Models (LLMs). Specifically, we utilized BERTopic \cite{grootendorst2022bertopic} to derive initial latent topics and Word2Vec \cite{Mikolov13Distributed, Mikolov13Efficient} to identify related keywords. This deterministic pipeline allowed us to explain the system's logic clearly to participants, mitigating the ``black box'' effect often cited as a barrier to AI adoption \cite{davison2024ethics}.} Users can either select the generated themes or manually add their own, as illustrated in Figure~\ref{fig:Automated theme generation}. A visualization-based interface displays theme distributions over time, indicating where specific themes appear most frequently in the transcript, as shown in Figure~\ref{fig:theme_viz}.

\newpage
\small 
\begin{longtable}{|c|c|c|p{1.3cm}|p{1.1cm}|p{3.6cm}|p{3.4cm}|}
\hline
\textbf{Participants} & \textbf{Age} & \revadd{ \textbf{Gender}} & \textbf{Years of Experience} & \textbf{\revised{Highest Degree}} & \textbf{Research Area} & \revadd{\textbf{Tools used for QDA}} \\
\hline
P1 & 26 & \revadd{F} & 4 & \revised{PhD student} & Human-AI collaboration, computing education accessibility &  \revadd{Google Docs, Google Sheets, Miro} \\
\hline
P2 & 27 & \revadd{F} & 2.5 & \revised{PhD student} & Design, culture, and accessibility &  \revadd{Google Docs, Google Sheets, Miro} \\
\hline
P3 & 24 & \revadd{F} & 3 & Masters & Accessibility &  \revadd{Google Docs, Google Sheets, Zoom, \revadd{FigJam}} \\
\hline
P4 & 33 & \revadd{F} & 7 & PhD & User studies on AI technology &  \revadd{Zoom, Webex} \\
\hline
P5 & 27 & \revadd{F} & 5 & \revised{Masters student} & Human-Computer Interaction (HCI) &  \revadd{Word, Miro, Microsoft Teams, FigJam, ChatGPT, iOS voice recording app, Figma} \\
\hline
P6 & 26 & \revadd{F} & 4 & \revised{Masters student} & Language learning, VR, ASR experiences &  \revadd{Google Sheets, Zoom, Atlas.ti, Otter AI, Whiteboarding tools, AirTable, Descript} \\
\hline
P7 & 31 & \revadd{F} & 6 & \revised{PhD student} & Chatbots, job support technologies &  \revadd{Excel, Miro, Microsoft Teams, Otter AI} \\
\hline
P8 & 40 & \revadd{M} & 2 & \revised{PhD student} & Captioning for d/Deaf and HoH users &  \revadd{Google Docs, Google Sheets, Miro} \\
\hline
P9 & 32 & \revadd{F} & 6 & \revised{PhD student} & Human-Computer Interaction &  \revadd{Google Docs, Zoom, Whisper, NVivo, Atlas.ti, Open source QDA on Github} \\
\hline
P10 & 26 & \revadd{M} & 3 & \revised{Masters student} & AI in clinical decision-making, model analysis &  \revadd{Google Docs, Google Sheets, Otter AI, Descript, Deedose} \\
\hline
P11 & 28 & \revadd{F} & 2 & Masters & Web and cognitive accessibility, digital tech use & \revadd{Google Docs, Excel, Microsoft Teams, Deedose} \\
\hline
P12 & 33 & \revadd{M} & 6 & \revised{PhD student} & AI and media forensics, public sentiment & \revadd{Google Docs, ChatGPT, Gemini advanced, Claude} \\
\hline
P13 & 35 & \revadd{F} & 3 & Masters & Early childhood and home education & \revadd{Excel, Microsoft Teams, Deedose, Covidence} \\
\hline
P14 & 26 & \revadd{F} & 3 & \revised{PhD student} & Deepfake detection tool design & \revadd{Google Docs, RefAI} \\
\hline
P15 & 23 & \revadd{F} & 3 & \revised{PhD student} & Accessibility for older adults and neurodivergent users & \revadd{Google Docs, Excel, Miro, Zoom, Taguette} \\
\hline
P16 & 29 & \revadd{F} & 3 & Masters & HIV services and relationship dynamics & \revadd{NVivo} \\
\hline
\caption{Overview of participant demographics, academic background, research expertise, and prior experience with qualitative data analysis (QDA) tools. The table includes information on age, \revadd{gender}, years of experience, highest degree attained, research domain, and QDA tools used by each participant. \revadd{Gender is denoted as F (Female) and M (Male).}}
\label{tab:participant_info}
\end{longtable}

\subsection{Procedure}
We conducted a formative study involving a pre-study demographic survey, an exploration of ChromaScribe, a semi-structured interview, and a post-study survey. The study was reviewed and approved by the university’s Institutional Review Board (IRB). 
The pre-study survey contained questions on demographic details, QDA experience, coding approaches, and the use of multimodal data. Participants were screened based on their research experience and inclusion criteria.
\revdel{Each interview began with a consent process and included questions on participants' current QDA workflow, QDA tool use and preferences, and current challenges related to the QDA process and tools. }
\revadd{Each interview began with a consent process and included questions on participants' current QDA workflow, QDA tool use and preferences, and current challenges related to the QDA process and tools.} In the next phase, participants were introduced to ChromaScribe. After watching a demo video, participants were given time to explore the webtool independently and then asked to complete four structured tasks including (a) locating a participant who mentioned a selected theme at least three times and (b) playing the audio of a segment where the interviewer mentioned a selected theme.
These tasks were designed to prompt participants to explore how the tool supports theme discovery\revadd{, navigation, transcript-audio alignment, search} \revdel{and navigation} \revadd{and to ensure that all participants engaged with the core functionalities of the prototype.} Participants were informed that the prototype is preliminary and only meant to facilitate the discussion.

\revadd{To address RQ1, participants were introduced to three distinct types (i.e. degrees of delegation) of AI involvement in qualitative coding: human-only coding, AI-initiated coding, and human-initiated coding, after the completion of four tasks on ChromaScribe.}  
The three coding types were presented to participants in a written format to elicit preferences and perceptions during the interview. Each participant was given a brief textual description of the three types and allotted two minutes to read them carefully. They were then asked to rank the coding types from most to least preferred. We further prompted them to share the reason for their preference.
\revadd{To minimize bias toward any particular AI delegation paradigm, the prototype was introduced prior to presenting participants with the three AI involvement modes and the participants were encouraged to explore different workflows as they desired.}
\revadd{To address RQ2,} we then asked about their collaborative practices in QDA, their thoughts on subjectivity and bias in the process, and how QDA tools could address bias. Lastly, \revadd{to address RQ3, participants were asked about their current reliance on transcripts, audio, and video during analysis and} we asked about their use \revdelII{of} and thoughts on multimodal data in QDA.

Two pilot interviews were conducted in person to refine interview questions and timing. Following this, an in-person focus group with two participants was held in a research lab at a university. Due to scheduling challenges, the remaining 14 interviews were conducted individually via Zoom, lasting approximately 90 minutes each. As an incentive, participants received a \$30 gift card upon completing the post-study survey.

\subsection{Data Analysis}
All interview sessions were recorded, with individual sessions transcribed through Zoom’s transcription feature and the focus group session transcribed through Otter.AI. Transcripts were then manually cleaned using Google Docs and coded with Atlas.ti. Data were analyzed using axial coding, a process of relating codes (categories and concepts) to each other, often involving a combination of inductive and deductive thinking to develop more abstract conceptual categories from initial codes \cite{Strauss1990Basics}. Two researchers independently conducted word-level and sentence-level coding to identify initial codes. 
\revised{To develop the initial codebook, both researchers independently coded the same three transcripts, then met to compare their coding decisions, resolve discrepancies, and refine a shared set of codes. The resulting codebook, maintained in a spreadsheet format, detailed codes, thematic descriptions, and relevant examples. Once finalized, one researcher applied the codebook to the remaining transcripts. When new codes emerged during this phase, they were documented and discussed with the second researcher to determine whether they represented new concepts or could be integrated into existing codes, with the codebook updated accordingly. This iterative process ensured that coding remained consistent and reflexive across the dataset.} Key themes included participants’ sense of ownership in the QDA process, anticipated benefits and concerns of incorporating AI \revchgII{in}{into} QDA, and how it compares with collaborative efforts with other researchers. 
\section{Results}

\textit{RQ1. \revchg{Which degree of AI involvement (human-only, human-initiated, AI-initiated) in QDA is most preferred by qualitative researchers, and what are the factors leading to this preference?}{How do qualitative researchers’ preferences for different degrees of AI involvement in QDA (human-only, human-initiated, and AI-initiated coding) vary across different contexts, and what factors shape these preferences?}}
\begin{table}[h!]
\centering
\begin{tabular}{|p{2.5cm}|p{5.7cm}|p{5.7cm}|}
\hline
\textbf{Coding Method}       & \textbf{Most Preferred} & \textbf{Least Preferred} \\ \hline

Human-only \newline Coding           & (n = 7) \newline - Comprehensive understanding of the data  \newline - Complete control over the process and higher reliability.& (n = 5) \newline - Requires significant time and effort from the researcher. \\ \hline
AI-Initiated Coding          & (n = 5) \newline - Reduced time for repetitive coding tasks and minimized mental effort.
\newline - Frees time for higher-level analysis and decision-making.& (n = 4) \newline - Less control over the coding process leading to less reliable results. \newline - AI generated themes may lack depth and context
\\ \hline
Human-Initiated Coding       & (n = 4) \newline - Leverages researchers’ contextual understanding of study domain and setup. \newline - AI supports verifying codes without compromising control of the process.& (n = 7) 
 \newline - Discomfort and distrust in AI verifying human QDA work.
 \newline - The additional layer of coding and verification increases researcher's workload\\ \hline
\end{tabular}
\caption{This table presents the number of participants who rated each method as the most preferred and the least preferred. It also lists the main factors that led to these preferences. To note, while we include the number of participants who preferred each coding method for transparency, our study was primarily qualitative in nature.}
\label{tab:coding_preferences}
\end{table}


\subsection{Human ownership and oversight in QDA}
\label{sec:Human_ownership_and_oversight}
A key factor influencing participants’ preference for human-AI collaboration styles was the sense of ownership each style gave. Many participants preferred human-only (n=7) and AI-initiated (n=5) approaches over the human-initiated approach, where AI verifies or revises human-generated codes. Several participants felt uncomfortable with the human-initiated approach, as they perceived a loss of authority or identity as a qualitative researcher. 
P5 voiced this discomfort, \textit{``Because I have to do everything initially... You [AI] are going to make me do it. You are going to verify... I mean, are you doubting me? It is that superiority complex that comes in. Are you doubting my work, like you were born yesterday, right?''}
Rather than viewing the human-initiated approach as productive collaboration, they felt that it undermined their expertise and professional autonomy. In their preferred (AI-initiated) approach, the human researcher was positioned to accept or reject AI-generated suggestions, not the other way around.

Interestingly, people who preferred the human-initiated approach also mentioned ownership as a primary motivation. As P2 explained: 

\begin{quote}
    I would rather myself go through it first and then, like, have AI check my work -- or I have not checked that, like I did not miss anything. Because, in a sense, that makes me feel like I own more of the results. That's what makes me a researcher. As a researcher, if I'm going through a PhD program and doing years of training, to be trained in this, I feel like I should have something to offer if it is an integral part of my job. Like, I'm happy to use AI as an assistant, but I do not think that it should [lead it].
\end{quote}

In both cases, ownership emerged as a guiding principle, whether it meant taking the first pass or making the final call. This emphasis on ownership contrasted with how participants viewed collaboration with human peers. 
Participants acknowledged the centrality of (human) collaboration in QDA, emphasizing the need for QDA tools that support collaborative analysis. As one participant (P10) noted, \textit{"I think collaboration is super important. like, we always need to have multiple coders working on it. But it's also important to know when coders disagree, and to see, like the history of that how the codes evolve over time."} Shared authorship in human collaboration was not only accepted but often valued; disagreement and negotiation were deemed natural in the QDA process where codes and themes are co-constructed through discussions and iterative refinement for a rigorous analysis. However, participants were more ambivalent when considering collaboration with AI in this process. While they were open to incorporating AI into their workflows, sharing interpretive authority with AI raised concerns about ownership and even impact on the professional identity of the researcher. 
\revadd{Across participants’ preferences for human-only, human-initiated, and AI-initiated coding, we observed a consistent pattern in how researchers conceptualized delegation. Participants were broadly willing to delegate low-effort, repetitive, or exploratory tasks such as scanning large datasets, identifying candidate codes, or surfacing recurring patterns, to AI systems but they resisted delegating interpretive authority, particularly for ambiguous, or context-sensitive coding decisions.}
Despite these reservations, most were open to workflows with asymmetrical collaboration that ensured that AI played a supporting role.
\revadd{While preferences regarding AI involvement varied across participants, researchers working in industry settings consistently expressed openness toward AI-supported qualitative analysis. Notably, all five industry participants, preferred workflows that involved AI to some degree, selecting either human-initiated or AI-initiated coding over fully human-only approaches. For example, P3, an industry researcher who regularly used AI features embedded in tools such as FigJam, favored AI-initiated workflows for tasks such as summarization and pattern detection, emphasizing the substantial time savings while retaining human oversight. Other industry participants preferred human-initiated approaches that positioned AI as a verification or support mechanism. P16, who worked with large, multi-country datasets, described AI as valuable for confirming themes and reducing human error when managing complex, distributed analyses. Together, these perspectives suggest that industry researchers were generally receptive to AI involvement in QDA, with variation centered not on whether AI should be used, but on how responsibility and control should be distributed between human researchers and AI systems.}
Participants consistently framed AI’s role as that of an assistant, rather than an equal peer or a ``co-pilot.'' In this assistant role, AI was seen as valuable for surfacing preliminary insights and improving efficiency, especially when managing and analyzing large datasets, provided that the human researcher remained the final decision-maker. P2 commented on AI’s expected efficiency, predicting that it could reduce task time from four hours to two.

\subsection{Mistrust in AI's Interpretive Capacity and Data Confidentiality}
While concerns about ownership and identity shaped attitudes toward AI collaboration, participants also expressed reservations based on trust, or rather mistrust, in AI’s interpretive ability. P8 explicitly expressed this mistrust, warning that \textit{``the more agency we give to AI, the less reliable the results are.''}

This low level of trust was evident in participants' views on the human-initiated approach, which was the least preferred option for seven out of sixteen participants. The approach was seen as inefficient as they would have to generate the initial codes, have the AI check the work, and re-verify the results themselves at the end. This perceived need to personally finalize and validate coding outcomes made the added layer of AI verification feel unnecessary or redundant. P6 noted that she would not feel comfortable relying on AI to validate her work, explaining, \textit{``I think I prefer the process of getting some kind of baseline.. And if I sort out the process, I just want to finish the process. So I do not tend to want to... start the process, and have the AI check it. Cause I wouldn't trust it as much, I guess.''}

One reason for this need to recheck the AI's work is that unlike human collaborators, AI was perceived as lacking interpretive depth and nuanced, contextual understanding that researchers bring to complex qualitative data based on their expertise and disciplinary training. As a result, participants raised concerns about the rigor of the analytical process, noting that AI would produce surface-level themes that failed to capture deeper meanings. Several participants noted that AI's inability to grasp the broader context of a study made it difficult to trust its coding decisions. Interestingly, the second reason for disliking the human-initiated approach was not about AI's performance, but its perceived role in the process; participants simply did not want AI to take the role of a supervisor. As participant (P7) explained, \textit{``So when ending in a human making the decision, checking the AI's work is really important. Which is why human initiating coding - I do not think I would ever do that. Because I want to check the AI's work. I do not want the AI to check my work.''}

Data confidentiality also emerged as a crucial concern, profoundly influencing how participants selected QDA tools and whether they even used them, especially those involving AI. For instance, P12 stated that his primary reservation in using QDA tools was privacy, and his willingness to use AI tools hinged on the availability of privacy-preserving features: \textit{``My only issue would be privacy… All of the tools [Claude, Gemini, ChatGPT] provide privacy mode. So you don’t give them the content you are analyzing.''} This sentiment was echoed by P14, who expressed a clear preference for local data processing to prevent sensitive information from being transmitted to third parties: \textit{``If this tool is just local, it’s on my desk. It’s not… giving out to a 3rd party, then give me results back, that’s fine. I want something like very like localized.''} 

This fundamental concern led participants to articulate specific preferences and strategies for interacting with AI tools to safeguard their data.
Beyond relying on QDA tool features, some participants described proactive measures to de-identify data before AI interaction. P5, while acknowledging the utility of tools like ChatGPT for thematic analysis, detailed a cautious approach: \textit{``Not saying that I literally put everything in ChatGPT as it is, like the whole transcribing file. No. Had to like still mask a lot of details, remove a lot of details.''} Similarly, P11 highlighted the risks associated with audio data containing personally identifiable information and suggested modifications: 
\begin{quote}
    For example, the audio that you're saving as a recording, it becomes a PI issue – like personal identifiable information. Because if you're storing the audio in its exact like, user tone [...] it's possible for someone to identify who might be speaking based on the context and the voice. So if I were to store audio, I would like disrupt the … or like voice, I'll make it into a more like robot... Or if I'm storing the voice, then I would make sure the tool has like all the encryptions and privacy protection stuff.
\end{quote}

Furthermore, overarching institutional guidelines played a significant role. P3, who works with sensitive data, underscored this by stating the limitations imposed by ethical review boards: \textit{``According to IRB, I cannot release my data of what participants said during the interview because [...] it might contain some kind of information which is private.''}  P16 also emphasized that \textit{``data privacy guidelines''} was a critical factor in selecting QDA tools, especially with AI involvement, stating, \textit{``We still do have a lot of data privacy guidelines on how our data is used and who has access to it, or what... on the Internet has access to it.''} The inability to verify what data is stored by AI tools, how it is processed, or whether it is used for training future models heightened their discomfort in using AI-based QDA tools.

These findings suggest that trust in AI for qualitative analysis is not merely about the accuracy of its outputs. It is tied to its perceived capacity to demonstrate contextual understanding and to operate in a way that respects data privacy and researcher control, standards many participants felt current AI systems did not yet meet.
\revadd{Participants did not converge on a single preferred degree of AI involvement. Instead, preferences reflected context-specific trade-offs shaped by researchers’ prior experience with qualitative methods, familiarity with AI tools, collaborative practices, and data sensitivity, illustrating how preferences for human-only, human-initiated, or AI-initiated coding are contingent on the analytical goals and constraints of a given research context. (see Section ~\ref{sec:Human_ownership_and_oversight}.)}

\subsection{Collaboration as a key to reduce bias}

\textit{{RQ 2. What are qualitative researchers’ perspectives on bias in qualitative data analysis and the use of AI for mitigating bias?}}

Although human-only coding was the most preferred method, some participants recognized the potential of AI-initiated coding as an additional collaborative layer for managing bias in qualitative research. All participants acknowledged the presence of cognitive bias in human-led qualitative analysis and that collaboration is key to mitigating bias. 

However, when asked whether QDA tools, with or without AI, could help reduce bias, participants expressed mixed views. Many participants were skeptical of AI’s current ability to address bias without human oversight. More specifically, they were divided on whether AI could act as a meaningful collaborator in this context. For most, addressing bias was seen as a deeply contextual and reflective process, which is fundamentally human and can only be provided by a human collaborator: 
\textit{"I think it's hard to like tease out the bias part. I think sometimes we just need a 3rd person to tell me this is the actual interpretation"} (P14). In their view, AI lacked the contextual sensitivity and critical judgment needed to challenge or validate human assumptions in a meaningful way.

Despite these reservations, participants remained open to the idea of using AI as a supplementary way of checking for bias, either for cases where human collaborators were unavailable or for enhancing current human collaboration. For example, P1 shared, \textit{``I use [ChatGPT] as like a check for my own analysis to see, `oh, they found this, but I did not find this' or `I identified more themes and they did more.' If I do not have a collaborator, I use ChatGPT as sort of like that extra check.''} 
Some participants suggested features that would enable QDA tools to better support bias reduction. For example, P3 noted that QDA tools could check for bias by treating transcripts equally, avoiding cherry picking data that could lead to confirmation bias. She further mentioned that QDA tools could help organize and structure data for a more impartial overview of the data, though noting that this alone is not sufficient to reduce bias. Others envisioned how initial theme suggestions or confirmation prompts could assist researchers in minimizing bias by aiding self-reflection. Notably, P5 mentioned that bias is not something that can be entirely engineered out of qualitative research: \textit{``the whole point of user interviews has to have that human touch. There is, you cannot deny there is going to be some amount of bias.''} 

\revaddII{Participants further highlighted how current human collaboration addresses the issue bias by making interpretations visible and open to challenge. Participants described aligning interpretations across teams, such as checking whether \textit{``the themes that we are finding here correlate with the things that they are finding there''} (P8) and resolving discrepancies through discussion, where \textit{``[we] combine our codes… [and we] may have discrepancy during the open coding process. There is a discussion in the process of merging the codes together, we'll be able to reach higher themes out, which is probably a more refined and in detail''} (P14). Rather than eliminating subjectivity, collaboration surfaced them, requiring researchers to critically examine their assumptions. As P12 noted, researchers must \textit{``separate yourself… and not defend your own work.''} Others emphasized that reflexivity also involves accountability for bias in data collection and analysis, such as recognizing and correcting errors: \textit{``this was a leading question… can I trim this?''} (P5). At a broader level, collaboration was framed as essential for interpretive rigor, where meaning is refined through ongoing discussion of what constitutes \textit{``the most rich information… [while] being true to what is being said''} (P16). These iterative conversations and shared workflows support the validation and refinement of interpretations.}

In all, while participants saw potential in AI to complement their efforts, bias reduction in qualitative analysis was seen as a human-driven process, benefiting most from collaboration and perspectives of multiple researchers. While QDA tools offer procedural support, they would be most effective when used alongside, rather than in place of, human collaboration.

\subsection{Integration of multimodal data in QDA}
\textit{{RQ3. How do qualitative researchers perceive the value of multimodal data (e.g., audio, video), and how do they incorporate it into their analysis practices?}}

The integration of multimodal data (e.g., text, audio, and video) into QDA offers researchers richer insights \cite{KuckartzRadiker2019, AlDhaleaEtAl2025}. This view was commonly shared by participants as well as they highlighted several advantages of incorporating multimodal data into their workflow for capturing information which is often lost in plain text.
To assess the \textit{actual} use of multimodal data in QDA, we asked participants to give an estimate of how much they relied on text (e.g., transcript, observation notes), audio, and video in percentages, adding up to 100\%.
Despite the recognized benefits of multimodal data, participant responses reveal a strong reliance on text-based transcripts as the primary medium for qualitative analysis. When considering how strongly they relied on text, estimates ranged from 40\% to 100\% (mean = 76\%). (To note, these estimates are intended to illustrate general trends in participants' use of different data types rather than to provide precise quantitative measures. Our study is primarily qualitative in nature.) Each participant indicated that text transcripts are essential for their analysis, with some (n=6) relying solely on text for coding and review. For example, P16 stated she uses audio only to gather insights on how participants express themselves but otherwise rely entirely on text transcripts. P15 and P9 both emphasized going through transcripts multiple times to ensure thoroughness in coding, suggesting that transcripts are manageable and useful for repetitive review.

Although most participants collected audio data in their studies, audio data usage in QDA averaged 60\% (SD = 11\%) among participants. 
Audio data is mainly used to identify change in tone and pitch, especially for detecting nuances in participants’ expressions, such as laughter or hesitation. P9 explained, \textit{``I used audio to identify laughter and tone''}, while P8 mentioned using audio for observing mood shifts during the interview. For some, audio was the only medium for data collection (i.e., no video recording, photographs, etc.). P16, engaged in a large-scale qualitative study across five countries, noted, \textit{``We strictly do audio recording... and then we transcribe those audio recordings and use the transcripts for our analysis.''} 

Video data was used even less frequently (mean= 7\%, SD = 8.9\%), though valued for capturing non-verbal cues and contextual information. P8 highlighted video’s role in \textit{``tracking mood changes''} and its importance in \textit{``interpreting gestures in American Sign Language transcriptions.''} 
Similarly, P3 found video indispensable, stating, \textit{``Video has been critical... I also transcribe what a person did and what they said,''} thereby integrating non-verbal and verbal cues directly into the textual analysis. Other people used video for gathering behavioral data. For example, P6 described using recordings to \textit{``review what people were doing or how people were interacting.''}
For usability testing, P15 mentioned, \textit{``We would do a video recording... we’d track how much time the user took, where they went...,''} capturing expressions like surprise visually. \revadd{P15's workflow closely mirrors the behavior-centric analysis seen in tools like CoUX \cite{Soure2022}, where visual evidence is essential for explaining user actions. However, for the majority of our participants conducting standard interviews, the text transcript effectively displaced the need for visual verification, leading to the ``flattening'' of data described earlier.} Video also played a role in verification, with P5 sharing, \textit{``It's just much easier when you want to \revchg{cross check}{cross-check} on some information.''} 

Researchers described various methods for aligning and navigating through different modalities such as transcripts, observation notes, audio, and video. The primary method was manually adding non-verbal information from audio or video files into the transcript. For example, P12 described incorporating observations from video or audio into notes during scenario-based studies. Another method was using a separate audio or video player to alternate between audio/video data and transcript. P11 illustrated \textit{``When I had to run my data analysis, it had to be both in conjunction with, like the team's transcript that we generated out of what they were speaking, and then the screen that we recorded. So I had to use these two side by side for understanding the context of the transcript along with the screen recording."} She did this in conjunction with the first method as she \textit{``used Excel to document quotes I saw in the screen recording and then correlate it with the transcript.''} While less common, P12 also mentioned occasionally using AI tools to extract themes from multimodal data, hinting at emerging practices.
Participants ideally wanted a better way of transitioning across different multimodal data, such as a synchronized view of video and transcript data. P8, a researcher working on American Sign Language (ASL) projects, primarily relied on text transcripts for data analysis. He shared that \textit{``If you are working with ASL transcriptions, I think that the video is definitely useful. It is like an important feature if you want to double check the transcriptions."} However, the effort required to review it during analysis was a significant barrier. He confessed, \textit{``If it was a streamlined process, where it is very easy to look at a video at a given moment or an audio file, maybe I would do it more. But it just takes so much effort to do it right now that I usually do not bother."}

The main reason behind underuse of multimodal data despite the recognized value was the time-consuming nature of engaging with non-textual media. The labor-intensive nature of working with these formats, particularly the difficulty of efficiently locating relevant segments, hindered participants from fully leveraging these sources. \revadd{This finding illuminates a key difference between our study context and prior work on multimodal systems. While CoUX \cite{Soure2022} addresses the technical challenge of synchronizing streams to analyze \textit{behavior}, our results show that in \textit{dialogue-centric} analysis, the primary barrier is cognitive and logistical: the text is ``good enough'' to analyze, so researchers actively avoid the friction of accessing the ``richer'' audio. Whereas analyzing the video is mandatory in usability testing, qualitative analysis of interview typically relies on the transcript, which acts as a convenient but reductive proxy.} P10 pointed out logistical hurdles, stating, \textit{``We don't usually analyze the video as much... it's harder than we just normally use the audio transcript.''} P14 elaborated on a situation where she remembered a participant saying something important but had to manually sift through multiple long transcripts or re-listen to audio to find it. She remarked that \textit{``sometimes it's hard to go back and locate what someone said... I wish I could reverse search across transcripts''} and expressed a desire for tools that could \textit{``pull out quotes I’m looking for without me rereading everything.''} This underscores a need for media-integrated systems capable of retrieving relevant moments in the desired modality (i.e., text or audio) based on semantic or contextual information, not just keywords.

\section{Discussion}
Our study revealed a growing openness among researchers to incorporate AI in the QDA process, diverging from earlier work where participants expressed significant resistance to AI involvement \cite{jiang2021, evers2018current, smith2018closing, kirsten2024, Gao2023CoAIcoder}. 
Our participants appreciated AI’s potential to streamline tasks in the initial stages of coding of QDA, such as grouping data and identifying common patterns for large qualitative datasets. They envisioned that this approach would allow them to focus on the deeper, interpretive aspects of data analysis that they find most engaging. 
Yet, significant concerns around ownership, trust, and potential biases made them hesitant about adopting AI in QDA. \revaddII{Crucially, we must interpret these preferences through the methodological lens of our study. The introduction of the prototype prior to evaluating delegation modes could have functioned as a framing device. However, this specific framing provided a valuable lens as exposure to AI-generated themes successfully surfaced participants' deepest anxieties around interpretive labor and yielded nuanced insights expected from qualitative research.} This discussion will explore these key themes, drawing on our findings and existing literature to propose directions for future work aimed at designing and presenting AI as an assistant, especially as a method of enhancing human collaboration \revadd{and shared sense-making} in QDA. 

\subsection{Navigating ownership and authorship}
Our participants emphasized the importance of maintaining ownership over the interpretive process, which influenced their coding preferences. Among the three methods presented, human-only coding was the most preferred because it gave researchers the strongest sense of control and interpretive depth. Although AI-generated suggestions were considered useful, participants underscored the importance of staying connected with the data and retaining the final say. 
This is consistent with Jiang et al.'s findings \cite{jiang2021}, where participants emphasized that direct engagement with data\revchg{, despite its messiness, was vital for developing meaningful understanding and fostering serendipitous insights. They cautioned that handing over analysis entirely to AI, or allowing it to overly direct the process, would disconnect them from essential sense-making, thereby diminishing their emotional connection and potentially undermining the unique, researcher-driven aspect of qualitative discovery.}{ is vital for sense-making and serendipitous insight. They cautioned that over delegating analysis to AI could disconnect researchers from essential sense-making, weakening emotional connection and the researcher-driven nature of qualitative discovery.} Therefore, Jiang et al. advocated for a methodology where AI offers code suggestions, leaving the researcher to accept or decline them. Our results further revealed that direct connection to the data also influenced their identity as a qualitative researcher. For instance, P1 shared that performing data analysis herself affirmed her identity as an HCI researcher and gave her a deeper sense of ownership over the findings. 

While the findings mainly focused on individual researcher's QDA workflow, the introduction of AI could reshape collaborative workflow and ownership as well. First, AI could reshape collaboration by enabling more asynchronous workflows. Because collaborators can independently review and respond to AI-generated inputs, the perceived need for real-time discussion may decrease. This can support scalability by allowing larger teams to divide analytic work into smaller tasks with system support the division aspect as well. At the same time, reduced synchronous interaction risks weakening collaborative sense-making, as there would be fewer shared moments for deep, dynamic discussions.  Next, the \revchgII{AI-generated coding approach decrease the negotiation phase during the collaborative coding as collaborators shift their discussion from \textit{``What do we think this means?''}to \textit{“Do we accept or reject the AI’s suggestion?”}}{indiscriminate integration of AI risks fundamentally reconfiguring the interpretive labor of qualitative coding. Our findings show that the current analysis workflow is inherently collaborative and reflexive, driven by moments of ``productive friction" where researchers question, justify, and reconcile interpretations to produce rigorous findings. However, the introduction of AI-suggested codes could shift the workflow to center around evaluating system-generated suggestions (\textit{“Do we accept or reject the AI’s suggestion?”}), replacing co-construction with reaching consensus. Consequently, the meaning of ownership could be altered as the act of approving rather than authoring interpretations, and the QDA results may no longer be deeply rooted in human perspective.}

Given the importance of the interpretive process, designers of AI-assisted QDA should be careful of these potential impact of AI. First, to support collaborative sense-making rather than shortcut it, AI-generated codes should be positioned as \textit{artifacts for discussion}, not final suggestions to be individually approved or rejected. For example, \textbf{the system could require multiple collaborators to independently evaluate codes that the system is less certain about, and for codes that receive different evaluations, the system could explicitly prompt researchers to articulate why they accept or reject a suggestion.} Since the researchers are not critiquing each other’s idea but a third party’s idea, it might also lead to more open discussions. This might especially be helpful to less-experienced researchers, who have reported feeling less confident in the process and being more open to AI. 

Future interface designs could support this desire for greater researcher ownership and transparency and facilitate robust collaborative sense-making by incorporating features that clarify the origins and evolution of codes. One option is a split-screen layout where researchers can view AI-suggested codes on one side and their own manual codes on the other, enabling easy comparison and informed decision-making. Another option is to \textbf{tag the origin of each code, such as ``AI-generated," the name of the researcher who coded it, or “researcher-approved” for AI-generated code that has been human-approved.} This approach could also be used to show the level of agreement among researchers (e.g., ``Below 50\% agreement" tag) to highlight codes that still need further discussion and distinguish collaboratively refined codes. Beyond transparency, such features enable AI to actively support delegation by signaling where human involvement is required and can flag analytically difficult segments for discussion, allowing AI to offload routine coding while directing human attention to unresolved interpretations.

This would address P10’s interest in QDA tools that let the user know when coders disagree. He further called for tools that show how codes evolved over time. For this purpose, we can draw inspiration from HistoryFlow \cite{viegas2004studying}, which visualizes human dynamics in group editing on Wikipedia to make authorship transparent. \textbf{Similar visualization could be used in QDA tools, through which researchers can see how conflicting codes were resolved and how their input shaped the final results.} This would not only be helpful for an individual researcher, but also for fostering shared accountability and enabling teams to trace the provenance of collective interpretations, a key consideration in CSCW research. \revaddII{Recent work by Ye et al., which focused on how AI could be incorporated in the collaborative workflow without harming the deep reflections, took a similar approach by displaying analysis history as well as providing in-situ reflexive prompts and scaffolding collaborative interpretation \cite{ye2026reflexis}.} 
Future work could explore whether this type of visualization affects researchers’ perceived control, collaborative efforts, and willingness to collaborate with AI systems. \revaddII{Furthermore, future research can examine how specific collaboration structures, such as power dynamics between lead researchers and research assistants, or interdisciplinary team configurations, shape these delegation preferences and workflows.}

\subsection{Establishing Trust: Performance, Process, and Purpose}
Along with ownership, lack of trust in AI-generated codes was one of the main reasons limiting participants’ willingness to adopt AI-assisted QDA tools. 
Lee and See's framework \cite{lee2004trust} provides a useful lens to understand the issue of trust. They operationalize trust as a combination of performance (i.e., reliability in achieving goals), process (i.e., clarity of inner workings), and purpose (i.e., alignment with user goals). We use this framework to organize our discussion and future work directions.

\subsubsection{Performance}
Participants assessed the performance of AI tools in QDA based on efficiency and reliability. Participants’ openness to AI as a supportive assistant was mainly due to anticipated efficiency, as they saw potential for such tools to automate or streamline repetitive tasks such as initial coding. They noted that providing AI with clear prompts and rules, combined with its high computational capacity, could simplify researchers’ roles to verifying outputs. On the other hand, people were largely skeptical of the reliability of the results due to the perception that the AI would not be able to capture nuanced, context-specific data accurately. This skepticism aligns with broader concerns around AI’s inability to fully grasp human meaning-making, raising doubts about the quality and credibility of its outputs.

Participants' mixed trust in AI performance reflects findings in prior studies, which noted AI's strengths in improving coding quality but questioned its effectiveness in handling complex interpretive tasks \cite{gao2023}. In prior work by Zhang et al. in the broader domain of AI-assisted decision making, the researchers explored two approaches of increasing trust: displaying confidence level and explaining the decision model \cite{Zhang2020}. Their findings indicated that users were more inclined to accept AI recommendations when higher confidence scores were shown, while model explanations did not have a significant impact on people's acceptance. This contrasts our participants' desire to see model explanations behind the codes that the AI has applied. It is possible that the ideal approach for enhancing trust differs by domains, and that in the context of QDA, model explanations may play a more critical role in building trust. Future work could explore this direction by comparing approaches such as showing confidence scores (i.e., correctness likelihood) when assigning codes, model explanation, or a list of alternative codes, on researchers' trust and willingness to integrate AI into their workflow. Additionally, integrating mechanisms that allow researchers to customize the extent of AI involvement based on the dataset size or complexity might foster interest in using AI in QDA by giving them more agency.

\subsubsection{Process}
In prior work, researchers emphasized the importance of transparency in AI’s coding processes and advocated for tools that can support marginalized populations while maintaining ethical integrity \cite{willson2019,lake2019}. If these tools reinforce dominant narratives or exclude outlier voices, they risk replicating harmful assumptions. Our results further confirmed this need for transparency.
A potential way of making the AI analysis process more transparent is through clear communication about the process, such as \textbf{explicit explanations of code suggestions (e.g., ``This excerpt was coded as \textit{`lack of trust in AI'} as the participant expressed skepticism about AI's ability to understand context.") while highlighting the relevant words or phrases in the transcript.} This would make the inner workings more transparent, allowing the user to identify and revise questionable AI outputs. This transparency is not only beneficial for incorporating AI, but also for \revadd{supporting coordination and} collaboration between researchers. More specifically, future systems could similarly enhance collaboration by \textbf{visualizing coding discrepancies between researchers and offering AI-generated interpretations of the differences} (e.g., ``Based on the sections coded as \textit{`trust in AI'}, it seems like Researcher 1 interpreted the code as including both trust \textit{and} mistrust in AI while Researcher 2 focused exclusively on sections showing trust, excluding those on mistrust.") \revadd{From a delegation standpoint, our findings suggest that AI-assisted QDA should support how analytic work is divided between AI systems and human researchers while expecting systems to \textbf{explicitly signal segments that require human sense-making, discussion, or judgment.} In this framing, delegation is not about replacing interpretation, but about directing human attention to analytically difficult moments where meaning must be negotiated.}

\subsubsection{Purpose}
Lastly, it is important to ensure that the system aligns with the user's goals. Our interviews revealed that for many QDA researchers, these goals include maintaining strict control over data, especially when working with sensitive or private data. A similar finding was reported by Yan et al. \cite{yan2024human}, who found people's skepticism toward ChatGPT due to doubts about data security. However, current QDA tools often provide limited information on where and how data is stored and processed. This lack of transparency even led some to avoid QDA tools altogether, resorting to manual coding or general-purpose software (e.g., MS Word) where local storage is ensured. 
The desire for local storage of research data presents a new challenge for AI-based QDA tools as there is a trade-off between performance and security. Typically, earlier and smaller language models can run locally (e.g., GPT-2), providing users with greater control over their data, which makes them more suitable for handling sensitive information (e.g., in healthcare or vulnerable populations) \cite{arsturn2023}.
In contrast, newer and more advanced models like GPT-4 that offer enhanced performance are often cloud-based, making them difficult or impossible to run locally due to computational demands and proprietary restrictions \cite{openai2023gpt4}. This introduces potential risks related to storing and processing data on remote third-party servers, potentially exposing confidential research data to unauthorized access or breaches. As AI-assisted QDA tools continue to advance, decisions around model selection, data storage, and deployment infrastructure must be critically examined to balance performance with researchers’ ethical and privacy needs. At the very least, all related information should be readily accessible and comprehensive -- including a clear description of where and how the data is stored (e.g., on a HIPPA compliant cloud server) and how the data gets used (e.g., whether it will be used for LLM training) --  enabling researchers to make informed decisions based on their project requirements and institutional guidelines.

\subsection{Using AI for overcoming bias by enhancing human collaboration in QDA and leveraging multimodal data}   
Bias is an inherent challenge in QDA, whether introduced by human subjectivity or algorithmic limitations \cite{bierema2020cognitivebias, panch2019ai_bias}. Participants recognized the inevitability of cognitive bias in human-led qualitative analysis and largely agreed that collaboration, whether among humans or between humans and AI, is key to mitigating bias. Some were already using AI tools like ChatGPT to validate their manual codes, suggesting that AI can act as an additional layer of review to challenge assumptions.

While AI can be used to reduce cognitive bias, concerns about algorithmic bias was a related theme. Algorithmic bias reflects the systematic inaccuracies embedded in AI tools due to flawed training data, limited contextual understanding, or improper algorithm design \cite{panch2019ai_bias}. A prior finding uncovered the issue of algorithmic bias in QDA by noting ChatGPT’s lack of contextual understanding, which required researchers to upload literature reviews to provide background information \cite{yan2024human}. Our participants echoed this concern that AI system's failure to understand the contextual nuances of a dataset may produce biased outcomes that reflect an incomplete view of the data. For example, in a healthcare study, a participant might say, “I stopped taking my meds because I felt better,” which a general-purpose AI could misclassify as `positive health outcome.' However, a trained medical researcher would recognize and tag this as a `potential non-adherence issue,' an important red flag in chronic illness management due to factors such as mistrust in the prescription or socioeconomic barriers to continued use.

To help researchers detect and address bias, both their own and that of the AI, future QDA systems could incorporate features that encourage reflective practice. One approach is to ask the users to explicitly explain a sample of their codes and themes before the system attempts broader application of these themes to the rest of the transcript. Through this process, it would not only learn the contextual knowledge, but it could surface potential inconsistencies or bias and subtly prompt researchers to re-assess the appropriateness and neutrality of their coding. To combat confirmation bias, systems could also require users to enter their hypotheses and research questions \textit{before} the analysis process, and summarize the resulting codes (e.g., visualizing the distribution of codes across participants). This could help users evaluate and better understand how the results are aligned or misaligned with their initial beliefs while preventing unconscious bias from influencing the results. This not only aids in addressing bias but could also ensure that all collaborators are aligned on the project’s goals and interpretive framework.

Based on our results and prior work, we hypothesize that AI could also support mitigating bias in QDA by leveraging multimodal data. Previous work has shown that analyzing textual transcripts may be insufficient to convey the full meaning of interview data, yet multimodal data is underutilized in QDA \cite{craig2021engaging}. Our results further corroborate the limited use of multimodal data in qualitative analysis. This is a missed opportunity since audio and video recordings contain non-verbal cues (e.g., pauses, laughter, facial expressions) that offer additional insights even contradicting the spoken words at times. Overlooking these cues may reduce the quality of AI-generated codes as well. For example, an AI might misinterpret a participant’s sarcastic remark (e.g., “Oh great, another meeting”) as a positive sentiment, failing to recognize the tone of voice or facial expression that signals frustration.

We suggest future work to explore the integration of multimodal data within AI-driven QDA tools to help detect and summarize trends that would be too time-consuming for researchers to find manually. For example, as researchers scroll through interview transcripts, the system could highlight passages with emotionally significant cues (e.g., sustained pauses, changes in vocal pitch, or visible discomfort) using icons above the relevant phrases or modified background color. Hovering over a highlighted section could trigger a pop-up that displays relevant descriptions like “long pause,” “sarcastic tone,” or “averted gaze,” with playback options to hear or see the original moment. The system could also group and summarize recurring non-verbal trends (e.g., “increased vocal tension when discussing workplace policies”) across participants, helping researchers quickly spot patterns that might otherwise be missed in text-only analysis. This design would enable researchers to easily cross-reference the transcript with non-verbal cues, validate AI-generated codes, and/or navigate large datasets more efficiently by focusing on emotionally or behaviorally significant moments. This may reduce both cognitive and algorithmic bias by grounding analysis in a richer, more holistic view of participant responses.

\section{Limitations}
This study had several limitations that should be considered when interpreting the findings. First, \revadd{this study focuses on three degrees of AI involvement in QDA based on a single framework \cite{lubars2019delegability}, but there are broader and more nuanced ways of involving AI within qualitative research. For example, AI involvement could extend to analytical coaxing (i.e., creative prompting to perform deeper analysis) \cite{messner2025artificial}, reflections on researcher's high-level mental models \cite{gebreegziabher2023patat}, and supporting language comprehension \cite{yan2024human}. Next,} the gender distribution among participants was imbalanced, many were in technology-related field, and the sample size was relatively small, limiting the generalizability of the results. In addition, although the study included experienced researchers who have at least one year of experience in qualitative data analysis, their familiarity with QDA tools and AI tools varied, which may have affected their responses. \revadd{Since our results showed that the desired role of AI in QDA is dependent on the researcher's context, }future study is needed to verify if the findings hold for researchers with a broader range of experiences, backgrounds, and demographic distribution \revadd{in a setting where participants can flexibly negotiate AI roles throughout the analytic process.
Another limitation of our work is that} the discussion on the three degrees of delegation was based on textual descriptions rather than actual experience with a working tool. This hypothetical approach may not fully capture their experiences and perceptions in real-world interaction with such systems. \revaddII{Furthermore, we acknowledge that early exposure to the ChromaScribe prototype likely functioned as a framing device that anchored participants' mental models around AI-generated thematic suggestions. Had participants interacted with a different dynamic, such as an AI that strictly retrieves data segments based on human-defined queries without suggesting its own themes, their ranking of delegation modes and specific anxieties around interpretive labor might have shifted.} To note, while the study reports the number of participants who preferred each coding method for transparency, the research was primarily qualitative in nature. As such, the findings should be interpreted with an emphasis on thematic insights rather than statistical generalizability. \revaddII{Lastly, while our findings highlight that context shapes how researchers perceive and adopt AI in QDA, our study did not deeply examine how specific collaboration structures (e.g., solo work with supervision, distributed collaborations) influence delegation preferences. Future research should investigate how different collaboration scale (e.g., individual, small-team, and large-team) and norms (e.g., coordination practices, accountability mechanisms, and reflexivity) influence human-AI delegation in qualitative analysis.} \revdelII{In addition, future system development explore features that visualize the provenance of interpretation to support the collaborative transparency and trust issues identified in this study.}

\section{Conclusion}
Building on prior QDA studies that focused on AI’s effects on coding efficiency and reliability, our work revealed researchers’ desire for interpretive ownership, reassurance on data confidentiality, and support in collaborative coding and \revadd{shared sense-making} when considering integrating AI into their QDA workflow. \revchg{By applying Lubars and Tan’s delegation framework, we compared three interaction paradigms: human‑only, human‑initiated, and AI‑initiated coding.}{While we utilized Lubars and Tan’s delegation framework to compare three interaction paradigms (human‑only, human‑initiated, and AI‑initiated coding), our findings suggest that researchers view AI-supported QDA not as fixed delegation, but as an evolving, hybrid interaction.} The findings indicate openness to AI and a slight preference for AI-initiated coding over human-initiated coding, as it provides efficiency while preserving researchers’ \revchg{control over final decisions}{ownership and accountability over interpretive outcomes}. Participants also highlighted AI’s potential to reduce cognitive biases through complementary guidance rather than full automation, underscoring the value of \revchg{collaborative oversight}{transparent negotiation of meaning}. Finally, although textual data remains the dominant modality for QDA, researchers recognized the importance of multimodal inputs to capture richer emotional and contextual information and sought tools to better integrate multimodal data in their workflow. Based on these results, we propose future work directions for designing AI as an assistant \revadd{that align with the values of ownership, accountability, and shared sense-making} to support collaborative QDA.

\bibliographystyle{ACM-Reference-Format}
\bibliography{CSCW_References}

\end{document}